\documentstyle[twocolumn,aps,epsf,floats,prb]{revtex}
\begin{document}

\draft

\title{Strain-induced shift in the elastically soft direction of
epitaxially grown fcc metals}
\author{V. Ozoli\c{n}\v{s}, C. Wolverton, and Alex Zunger}
\address{National Renewable Energy Laboratory, Golden, CO 80401}
\date{October 20, 1997}
\maketitle

{\let\clearpage\relax
\twocolumn[
\widetext\leftskip=0.1075\textwidth \rightskip\leftskip
\begin{abstract}
The theory of epitaxial strain energy is extended beyond the harmonic
approximation to account for large film/substrate lattice mismatch.
We find that for fcc noble metals (i) directions $\langle 001 \rangle$
and $\langle 111 \rangle$ soften under {\it tensile\/} biaxial strain 
(unlike zincblende semiconductors) while (ii) $\langle 110 \rangle$
and $\langle 201 \rangle$ soften  under {\it compressive\/} biaxial
strain. Consequently, (iii) upon sufficient compression $\langle 201
\rangle$ becomes the 
softest direction (lowest elastic energy), but (iv) $\langle 110
\rangle$ is the hardest direction for large tensile strain. 
(v) The dramatic softening of $\langle 001 \rangle$ in fcc noble
metals upon biaxial tensile strain is caused by small fcc/bcc energy
differences for these materials.
These results can be used in selecting the substrate
orientation for effective epitaxial growth of pure elements and
$A_pB_q$ superlattices, as well as to explain the shapes of coherent
precipitates in phase separating alloys.
\end{abstract}
\pacs{PACS numbers: 62.20.Dc, 68.60.-p, 81.10.Aj}
]}

\narrowtext

When a material is compressed {\it hydrostatically,\/} its energy
rises steeply because all three crystal axes are deformed (dashed line
in Fig.~\protect\ref{fig:eepi}). When the same material is confined
coherently onto a substrate (``coherent {\it epitaxy\/}'') with
lattice constant $a_s$, the energy rises less steeply (solid line in
Fig.~\ref{fig:eepi}) since it is deformed only along the crystal axes
in the substrate plane and allowed to relax (and thus, lower its
energy) in the third direction $\widehat{G}$. This ``epitaxial
softening'' can be quantified by the dimensionless parameter
\begin{equation}
\label{eq:qdef}
q(a_s, \widehat{G}) = \frac{\Delta E^{\rm epi} (a_s,\widehat{G})}
{\Delta E^{\rm bulk} (a_s)},
\end{equation}
giving the ratio between the epitaxial increase in energy due to
biaxial deformation to $a_s$, and the hydrostatic increase in
energy due to triaxial deformation to the same $a_s$. Because the 
biaxial strain energy $\Delta E^{\rm epi} (a_s,\widehat{G})$ 
depends on the strain direction $\widehat{G}$, so does
$q(a_s,\widehat{G})$. In growing coherent epitaxial
films,\cite{epitaxy} it is desirable to minimize $\Delta E^{\rm
epi}(a_s,\widehat{G})$ [or, equivalently, for a fixed $a_s$ minimize
$q(a_s,\widehat{G})$], so as to avoid or reduce dislocations and
other strain-induced film/substrate defects.
It is hence important to select substrates $a_s$ and growth directions
$\widehat{G}$ that entail minimal strain energy. Harmonic continuum
elasticity
theory\cite{HB78,anast90,dmwood88,dmwood92,dblaks92,epi_review,marcus95,japanese}
makes  definitive predictions for the $a_s$- and
$\widehat{G}$-dependence of epitaxial strain energy:
(i) $q(a_s,\widehat{G})$ does not depend on $a_s$, and (ii) the
``softest direction'' is $\langle 001 \rangle$ if
$\Delta=[C_{44}-\frac{1}{2}(C_{11}-C_{12})]>0$, while if $\Delta<0$
then the softest direction is $\langle 111 \rangle$. For most fcc
metals and
semiconductors $\Delta>0$, whereas $\Delta<0$ for ionic salts
(PbS, AgBr, NaCl, KCl) and several bcc metals (Nb, V, Mo, Cr). 
The selection of substrate orientation
for many years has been guided by these
predictions of harmonic elasticity,
summarized compactly by the expression\cite{dblaks92,epi_review}
\begin{equation}
\label{eq:qharm}
q_{\rm harm} (\widehat{G}) = 1 - \frac{B}{C_{11} +
\Delta \; \gamma_{\rm harm} (\widehat{G})},
\end{equation}
where $B = \frac{1}{3} (C_{11} + 2 C_{12})$ is the bulk modulus and
$\gamma_{\rm harm} (\widehat{G})$ is a purely geometric function
of the spherical angles formed by $\widehat{G}$:
\begin{eqnarray}
\nonumber
\gamma_{\rm harm} (\phi, \theta) = 
\sin^2 (2\theta) + \sin^4 (\theta) \sin^2 \\
\label{eq:gamma_harm}
= \frac{4}{5} \sqrt{4\pi} [K_0(\phi,\theta) - \frac{2}{\sqrt{21}}
K_4(\phi,\theta) ].
\end{eqnarray}
$K_l(\phi,\theta)$ are Kubic harmonics\cite{kubic} (linear
combinations of spherical harmonics forming irreducible
representations of the cubic group), and the spherical angles
$\phi$ and $\theta$ are measured with respect to Cartesian axes
oriented along the edges of the conventional fcc cubic cell.
Figures~\ref{fig:qsexy}(b) and ~\ref{fig:qsexy}(c) show $q_{\rm
harm}(\widehat{G})$ of Au and Cu, demonstrating that
$\langle 001 \rangle$ and $\langle 111 \rangle$ are, 
respectively, the softest and hardest directions.

\begin{figure}
\epsfxsize=2.5in
\centerline{\epsffile{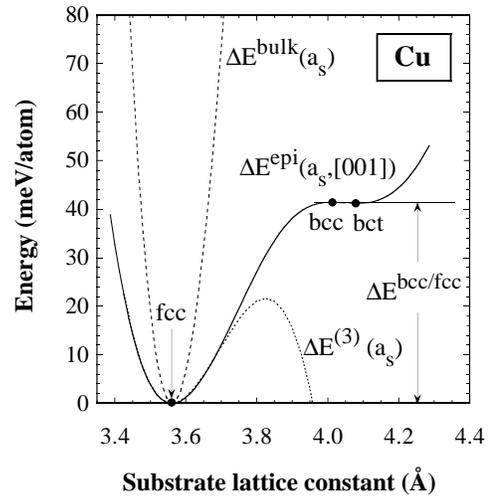}}
\caption{Epitaxial ($\Delta E^{\rm epi}(a_s,\widehat{G})$, solid line)
and hydrostatic ($\Delta E^{\rm bulk}(a_s)$, dashed line) deformation
energies for fcc Cu. The dotted line shows a fit to $\Delta E^{\rm
epi}(a_s,\widehat{G})$ using 3-rd order polynomial in $a_s$.}
\label{fig:eepi}
\end{figure}

\begin{figure}
\epsfxsize=3.0in
\centerline{\epsffile{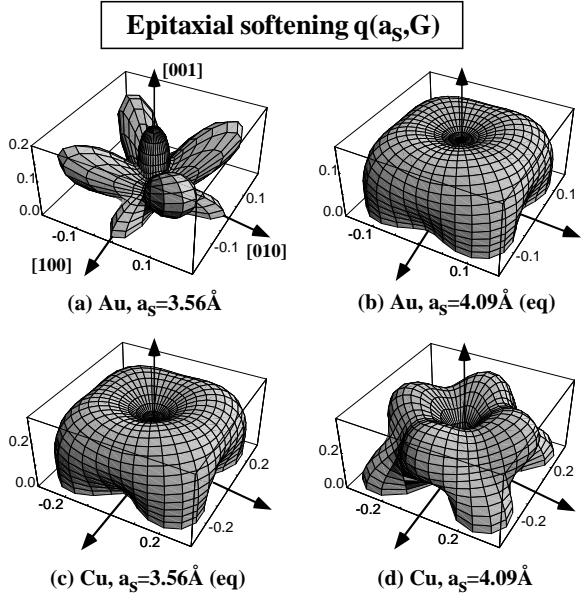}}
\vskip 3mm
\caption{Parametric plot of the epitaxial softening function
$q(a_s,\widehat{G})$ {\it vs.\/} substrate orientation $\widehat{G}$
for (a)--(b) Au and (c)--(d) Cu, at a few values of the substrate
lattice constant $a_s$. The equilibrium (eq) lattice constants of fcc
Au and fcc Cu are shown in (b) and (c), respectively.}
\label{fig:qsexy}
\end{figure}

One expects to find corrections to the harmonic behavior predicted by
Eq.~(\ref{eq:qharm}) due to the fact that strong deformations may
alter the electronic charge density (and thus, the elastic response)
beyond 
the validity of conventional harmonic elasticity theory. However, the
theory of elasticity itself does not indicate when and how
expression (\ref{eq:qharm}) will fail for a given system.
Figure~\ref{fig:eepi} shows that at least a fourth-order polynomial 
in $a_s$ is needed to reproduce the qualitative structure in 
$\Delta E^{\rm epi}(a_s,[001])$ for Cu, as obtained directly from
accurate electronic structure calculations (see below). The three
extremal points in $\Delta E^{\rm epi}(a_s,[001])$ suggest that a
simple extension of the harmonic elasticity to include third-order
elastic constants is not sufficient (the dotted line in
Fig.~\ref{fig:eepi} shows the behavior of the strain energy predicted
by a 3-rd order polynomial fit to a few data points around $a_{\rm
eq}=3.56$~\AA).
The goal of the present work is to propose a simple analytic formula
for $q(a_s,\widehat{G})$ [and therefore $\Delta E^{\rm epi}
(a_s,\widehat{G}$)] which works for arbitrary strain magnitude
and direction $\widehat{G}$.

We have calculated $\Delta E^{\rm epi}(a_s,\widehat{G})$ and
$q(a_s,\widehat{G})$ for Ag, Au, Cu and Ni along six principal
directions $\langle 001 \rangle$, $\langle 111 \rangle$, 
$\langle 110 \rangle$, $\langle 113 \rangle$, $\langle 201 \rangle$ 
and $\langle 221 \rangle$ using the local
density approximation (LDA),\cite{DF} as implemented by the linear
augmented plane wave (LAPW) method.\cite{lapw} This approach
gives the total energy as a function of biaxial
distortion, including all anharmonic effects. We
find that our results can be fitted accurately to a generalization of
the Kubic harmonic expansion of Eq.~(\ref{eq:gamma_harm}):
\begin{equation}
\label{eq:gammagen}
\gamma(a_s,\widehat{G}) = \gamma_{\rm harm} (\widehat{G}) + 
\sum_{l=0}^{l_{\rm max}} b_l(a_s) \, K_l (\widehat{G}),
\end{equation}
where $b_l(a_s)$ is now a function of $a_s$, and the angular momenta
extended beyond the harmonic limit $(l=0,4)$ to include
$l=6,8,10,\ldots$
The analytic representation of Eq.~(\ref{eq:gammagen}) allows us to
explore the elastically soft and elastically hard directions
for {\it any\/} strain, even outside the regime of harmonic
elasticity. We find that for {\it fcc\/} noble metals $q(a_s,[001])$ and
$q(a_s,[111])$ soften as $a_s$ expands (tensile strain) while $q(a_s,[110])$
and $q(a_s,[201])$ soften as $a_s$ contracts (compressive
strain). Consequently, upon sufficient compression 
$\langle 001 \rangle$ is no longer the softest direction, but rather
$\langle 201 \rangle$ is (e.g., Ag, Au). 
Also, upon sufficient expansion, $\langle 111 \rangle$ 
is no longer the hardest direction, which is now $\langle 110 \rangle$
(e.g., Cu, Ni). Finally, the
dramatic softening of $q(a_s,[001])$ upon biaxial expansion reflects the
existence of low-energy ``excited'' bcc and bct crystal forms. 
These results can better guide the selection of substrate orientation
for effective epitaxial growth.

\begin{figure}
\epsfxsize=3.2in
\centerline{\epsffile{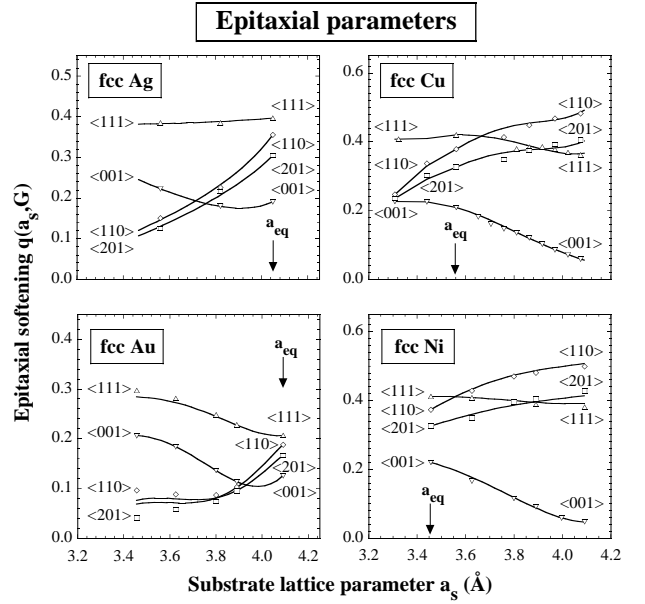}}
\caption{The calculated epitaxial softening functions
$q(a_s,\widehat{G})$ {\it vs.\/} substrate lattice parameter $a_s$ for
Ag, Au, Cu and Ni. Points represent the directly calculated LDA values
and lines show the fit using Eq.~(\protect\ref{eq:gammagen}).}
\label{fig:qs}
\end{figure}

We have performed LDA calculations for biaxial
compression ($a_s < a_{\rm eq}$) of Au and Ag, for biaxial expansion
($a_s > a_{\rm eq}$) of Ni, while for Cu we considered both expansion
and compression. Table~\ref{tab:Cij} gives the
calculated LDA 
elastic constants, which are typically 10-20\% higher than the
experimental values for Cu, Ag and Au.
Larger discrepancy in the case of fcc Ni 
is due to the neglect of spin-polarization effects in our LAPW
calculations. Figure~\ref{fig:qs} shows a line plot of 
$q(a_s,\widehat{G})$ {\it vs.\/} $a_s$ for a few directions
$\widehat{G}$, while Fig.~\ref{fig:qsexy} shows a parametric plot of
$q(a_s,\widehat{G})$ {\it vs.\/} $\widehat{G}$ for a few substrate
lattice constants $a_s$. We find several surprising predictions
relative to the 
expectations of harmonic elasticity. Although the harmonic expression
$q_{\rm harm}(\widehat{G})$ [Eq.~\ref{eq:qharm}] is exactly satisfied
at the equilibrium lattice constant $a_s = a_{\rm eq}$ [where
$\lim_{a_s \rightarrow a_{\rm eq}} b_l(a_s) = 0$ in
Eq.~(\ref{eq:gammagen})], this ceases to be so as $a_s$ deviates from
$a_{\rm eq}$. This is apparent from the dependence of $q$ on $a_s$
(absent in the harmonic theory), from the crossing of $q$ for different
$\widehat{G}$ values, and from the development of new lobes and minima
in Fig.~\ref{fig:qsexy} with the change in $a_s$. Such effects occur
at $\Delta a/a < 4\%$, suggesting a rather small range of validity of
the harmonic approximation. Furthermore, in the harmonic elasticity
theory, if $\langle 001 \rangle$ is the softest direction then 
$\langle 111 \rangle$ must be the
hardest direction, and vice versa. Figure~\ref{fig:qs} shows that this
does not hold for sufficiently deformed films: the hardest direction
in Ni and Cu for $a_s \gg a_{\rm eq}$ is $\langle 201 \rangle$, 
while the hardest directions for Ag and Au at $a_s \ll a_{\rm eq}$ are
$\langle 111 \rangle$ and $\langle 001 \rangle$.

\begin{table}
\caption{The calculated and experimental
(Ref.~\protect\onlinecite{huntington}) elastic constants of fcc Ag,
fcc Au, fcc Cu and fcc Ni (in Mbar).}
\par
\begin{tabular}{c|cc|cc|cc|cc}
& \multicolumn{2}{c|}{Ag} & \multicolumn{2}{c|}{Au} & 
\multicolumn{2}{c|}{Cu} & \multicolumn{2}{c}{Ni} \\
 & Calc. & Expt. & Calc. & Expt. & Calc. & Expt. & Calc. & Expt. \\
\tableline
$C_{11}$ & 1.52 & 1.24 & 2.14 & 1.86 & 2.30 & 1.68 & 3.30 & 2.47 \\
$C_{12}$ & 1.09 & 0.93 & 1.73 & 1.57 & 1.58 & 1.21 & 2.20 & 1.47 \\
$C_{44}$ & 0.61 & 0.46 & 0.37 & 0.42 & 0.99 & 0.75 & 1.36 & 1.25 \\
\end{tabular}
\label{tab:Cij}
\end{table}

The new results for fcc noble metals, apparent from our
self-consistent LDA calculations, are: 

(i) $q(a_s,[001])$ and, to a lesser extent, $q(a_s,[111])$ soften as $a_s$
{\it expands\/} (tensile biaxial strain). 

(ii) $q(a_s,[110])$ and $q(a_s,[201])$ soften as $a_s$ is {\it
compressed.\/}

(iii) As a result of (i) and (ii) above, we find that upon sufficient
compression, $\langle 001 \rangle$ is no longer the elastically 
softest direction, but $\langle 201 \rangle$ and, to a lesser extent, 
$\langle 110 \rangle$ are. The hardest direction upon
compression is still $\langle 111 \rangle$. 

(iv) Upon sufficient expansion, $\langle 111 \rangle$ is no longer 
the hardest direction, but $\langle 110 \rangle$ is (Cu, Ni). 
The softest direction upon expansion is still $\langle 001 \rangle$.

We find that result (i) is a reflection of the existence of a
{\it low-energy\/} bcc and bct ``excited'' structures.
Indeed, $\langle 001 \rangle$ strain applied to fcc lattice
defines a Bain path,\cite{bain} transforming fcc into bcc via
the body-centered tetragonal structure. This intermediate
structure is characterized by tetragonal unit cell dimensions $a$ and
$c$. When minimizing the total energy of a $\langle 001 \rangle$
biaxially deformed solid with respect to the out-of-plane lattice
vector $c$ at each in-plane lattice parameter $a_s$, one finds an 
``epitaxial $\langle 001 \rangle$ Bain path''\cite{alippi} 
(solid line in Fig.~\ref{fig:eepi}). For noble metals having the fcc
structure ($c/a=\sqrt{2}$) at equilibrium, this deformation path
contains the bcc ($c/a=1$) saddle point, and the bct ($c/a=0.96$
in the case of Cu) local minimum.\cite{alippi,kraft93,craievich94}
The low amplitude of the epitaxial $\langle 001 \rangle$ Bain path
relative to the hydrostatic path (Fig.~\ref{fig:eepi}) defines the
softness of $q(a_s,[001])$ via Eq.~(\ref{eq:qdef}), and therefore is a
direct manifestation of the small fcc/bcc and fcc/bct energy
differences. Indeed, the
epitaxial softening function at $c/a=1$ is given by
\begin{equation}
q(a_s,[001]) = \frac{\Delta E^{\rm bcc/fcc}}
{\Delta E^{\rm bulk}_{\rm fcc} (a_s)},
\end{equation}
where $\Delta E^{\rm bcc/fcc} = E^{\rm bcc}_{\rm tot} 
(V_{\rm eq}^{\rm bcc}) - E^{\rm fcc}_{\rm tot} (V_{\rm eq}^{\rm
fcc})$. Since in fcc noble metals 
$V_{\rm eq}^{\rm bcc} \approx V_{\rm eq}^{\rm fcc}$, the bcc
point is reached at $a_s \approx 2^{\frac{1}{3}} a_{\rm eq}$
(see Fig.~\ref{fig:qs}).

A similar argument explains the softening of $q(a_s,[111])$,
since the epitaxial $\langle 111 \rangle$ Bain path also 
connects cubic fcc ($c/a=\sqrt{6}$) and
bcc ($c/a=\sqrt{6}/4$) structures.\cite{kraft93} 
This property of the $\langle 001 \rangle$ and $\langle 111 \rangle$ 
Bain paths is caused by the cubic symmetry of the crystal at the bcc
point $c/a=1$, requiring it to be a local extremum of the total energy.
It is interesting that in zincblende solids (two interpenetrating fcc
lattices), tetragonal $\langle 001 \rangle$ 
expansion does not lead to an extremal point in 
$\Delta E^{\rm epi}(a_s,[001])$ since the crystal at the 
``bcc point'' $c/a=1$ does not posses higher symmetry than at 
$c/a \neq 1$. As a consequence, $\Delta E^{\rm epi}(a_s,[001])$ 
is a monotonically increasing function of $a_s$,
and $\langle 001 \rangle$ tensile strain does not lead to a softening
of $q(a_s,[001])$.

The consequences of our findings are as follows: (a) Films of 
fcc noble metals under {\it tensile\/} biaxial strain possess
the lowest strain energy for strain direction $\langle 001 \rangle$, 
while large {\it compressive\/} biaxial strains will have lower strain
energy for $\langle 201 \rangle$.
(b) When growing an $S_pL_q$ superlattice with
components having small ($S$) and large ($L$) lattice constants
(e.g., $S={\rm Cu}$, $L={\rm Au}$), the system rich in $L$ 
($q \gg p$) will have a low $\langle 001 \rangle$ 
elastic strain energy due to easy expansion of $S$, 
while a system rich in $S$ ($q \ll p$) will
have a low $\langle 201 \rangle$ elastic 
energy due to easy compression of $L$.

This work has been supported by the Office of Energy Research,
Basic Energy Sciences, Materials Science Division, U.S. Department of
Energy, under contract DE-AC36-83CH10093.

\end{document}